# AI-Powered Learning: Making Education Accessible, Affordable, and Achievable[1]

Ashok Goel, Design & Intelligence Laboratory, Georgia Institute of Technology; goel@cc.gatech.edu

*Abstract*— We have developed an AI-powered socio-technical system for making online learning in higher education more accessible, affordable and achievable. In particular, we have developed four novel and intertwined AI technologies: (1) VERA, a virtual experimentation research assistant for supporting inquiry-based learning of scientific knowledge, (2) Jill Watson Q&A, a virtual teaching assistant for answering questions based on educational documents including VERA's user reference guide, (3) Jill Watson SA, a virtual social agent that promotes online interactions, and (4) Agent Smith, that helps generate a Jill Watson Q&A agent for new documents such as class syllabi. The results are positive: (i) VERA enhances ecological knowledge and is freely available online; (ii) Jill Watson Q&A has been used by >4,000 students in >12 online classes and saved teachers >500 hours of work; (iii) Jill Q&A and Jill Watson SA promote learner engagement, interaction, and community; and (iv). Agent Smith helps generate Jill Watson Q&A for a new syllabus within ~25 hours. Put together, these innovative technologies help make online learning simultaneously more accessible (by making materials available online), affordable (by saving teachers' time), and achievable (by providing learning assistance and fostering student engagement).

## I. Problem Area

The emPrize project addresses two closely related grand challenges in learning and education. First, education is a "wicked" problem (Rittel & Webber 1973): it is open-ended and ill-defined, and it has multiple goals such as accessibility, affordability, achievability, and quality. Moreover, these goals often are in conflict with one another. For example, making teaching more effective (for instance, through individual human tutoring) typically makes it less affordable (by requiring more tutors). The emPrize project seeks to create and use AI technology for making quality education simultaneously more accessible, affordable and achievable.

Second, human learning is not only a cognitive process (Bruner 1960) but also a socio-cultural process (Vygotsky 1978): cognition and learning are situated in social and cultural contexts, and learning is socially mediated by knowledgeable others. Thus, while the use of AI for supporting individual tutoring for well-structured problems in closed domains can be quite productive, learning in general requires both social engagement and learning assistance. The emPrize project seeks to design socio-technical systems that include AI tutors but also engage social interaction and learning assistance.

Online learning in higher education manifests both of these challenges. For example, while massive open online courses (MOOCs) make higher education more accessible and affordable, often they also make it less achievable: MOOCs typically have lower completion ratios and lower student satisfaction than residential classes (Daniel 2012). Questions of the quality of learning in MOOCs persist as well. This is mostly because of lack of learning assistance and social interaction (Hollands & Tirthali 2013). (This is what achievability means here: the availability of learning assistance and social interaction so that the learner can accomplish her learning goals.). Thus, the specific goal of the emPrize project is to address the above two grand challenges in the context of online learning in higher education.

## II. Technical Solution

The emPrize project has developed an integrated and innovative set of AI technologies that combine knowledge-based AI and machine learning (ML) techniques as elements of a socio-technical system for online learning. In particular, we have developed four novel and intertwined AI technologies (1) VERA, a virtual research assistant that uses large-scale knowledge and knowledge-based AI for supporting inquiry-based learning of scientific knowledge, (2) Jill Watson Q&A, a virtual teaching assistant that combines knowledge-based AI and ML to answer learners' questions (3) Jill Watson SA, a virtual social agent that uses knowledge-based AI to promote learner interactions and community in conjunction with Jill Watson Q&A, and (4) Agent Smith that uses knowledge-based AI and ML to interactively train a Jill Watson Q&A agent for answering questions based on a new document such as a class syllabus or a user guide (such as the VERA users' guide).

### II.A. VERA

The VERA project addresses the issues of accessibility, achievability and quality of online education. Residential students in higher education have access to physical laboratories, where they conduct experiments and participate in research, thus discovering new knowledge grounding in empirical evidence and connecting it with prior knowledge. Online learners do not have access to physical laboratories, which impairs the quality of their learning. Thus, we have developed a virtual experimentation research assistant (called VERA for short) for inquiry-based learning of scientific knowledge (An et al. 2019a, 2019b): VERA helps learners build conceptual models of complex phenomena, evaluate them through simulation, and revise the models as needed. VERA's capability of evaluating a model by simulation provides formative assessment on the model; its support for the whole cycle of model construction, evaluation, and revision fosters self-regulated learning. Given that even

---

residential students have only limited access to physical laboratories, VERA is also useful for blended learning.

VERA is available online (http://vera.cc.gatech.edu) for free and public use. Although VERA is useful for any agent-based domain, for the specific domain of ecology, we have integrated it with Smithsonian Institution's Encyclopedia of Life that is available as an open source library and software (EOL; www.eol.org; Parr et al 2016). We have also created a set of video lessons on The Scientific Way of Thinking that rely on project-based learning using VERA; the video lessons are available on YouTube (https://youtu.be/Rn5e9fWmPqI) for free and public use. Finally, we have created a Jill Watson Q&A agent for assistance in using VERA; thus Jill too is now available for learning assistance in the wild.

*II.B. JILL WATSON Q&A*

The Jill Watson Q&A (Jill for short) project is part of our AI-powered socio-technical system: Jill seeks to support both the teacher and the learners in an online environment. On one hand, offloading answering routine questions to an AI teaching assistant can free up precious time for overloaded teachers. A teacher may use this time to reach more students, reducing the cost per student; alternatively the teacher may use the freed time to engage with students on deeper questions, enhancing the quality of learning. On the other hand, Jill provides learning assistance, and helps enhance learner engagement by answering their questions promptly, thus simultaneously making education more achievable.

In January 2014, Georgia Tech launched its Online Master of Science in Computer Science program (OMSCS; http://www.omscs.gatech.edu). The video lessons for an OMSCS course are delivered for free through Udacity. The online students interact with the professor and the teaching assistants on the Piazza web-based discussion forum. OMSCS currently has ~9,000 students and supplies ~9% of all MS in CS graduates in the country (Goodman, Melkers & Pallais 2019). Georgia Tech considers OMSCS to be very successful (Joyner, Isbell & Goel 2016; Joyner et al. 2019).

In Fall 2014, as part of the Georgia Tech OMSCS program, we created an online class on Knowledge-Based AI (KBAI; https://www.udacity.com/course/knowledge-based-ai-cognitive-systems--ud409; Goel & Joyner 2016, 2017). When we first taught the online KBAI class in Fall 2014, we found that the hundreds of students who took the class posted thousands of messages and hundreds of questions on the Piazza discussion forum for the class that overwhelmed the teaching staff. Thus, in 2016 we developed a virtual teaching assistant named Jill Watson that automatically answered routine, frequently asked questions posted on the online discussion forum (Goel & Polepeddi 2017, 2018); we will call this the Jill Watson Q&A (2016) agent. (The name Jill Watson derives from the use of IBM's Watson system in its initial construction.) Jill contained a digital library of Q&A pairs from previous semesters organized into question categories. Given a new question, Jill classified the question into a category, retrieved an associated answer, and returned the answer if its confidence value was high enough (>97%).

Since then we have built a new version of the Jill that answers questions based on class syllabi (instead of a databank of previous Q&A pairs). We have developed a novel ontology of class syllabi. We organize the new Jill's knowledge of the syllabus of a given class around this general ontology. We will call this the Jill Watson Q&A (2019) agent; from now on, whenever we mention Jill without any modifier, we are referring to Jill Watson Q&A (2019). Figure 1 illustrates the evolution of Jill.

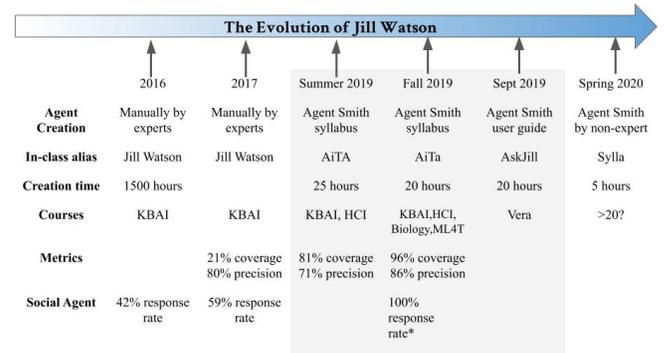

*Fall 2019 introduced a new Community-Building feature which is discussed further in Section II.C.

Fig. 1: The evolution of Jill Watson since its initiation in Spring 2016. We will describe the various entries in detail in different sections below. Briefly, in 2016, it took us ~1500 hours to create Jill; now Agent Smith helps create Jill in <25 hours. In 2016, Jill worked in for only one class; now Jill is working in six classes simultaneously. In 2017, Jill covered about 21% of question asked with a precision of ~80%; now Jill covers about 96% with a precision of ~86%.

Jill Watson Q&A (2019) uses an innovative 2-stage classification process. The first stage uses an ensemble of commercially available ML classifiers such as Watson, LEX, and AutoML. The second stage uses our own proprietary knowledge-based classifier. Thus, the 2-stage classifier combines knowledge-based AI and ML: it first uses ML to classify a sentence into general categories, and then knowledge-based AI to extract specific details based on the general classification. The answers derive from the knowledge of the syllabus.

All responses pass through a personality module that acts as an emotion modulator because we found that learners respond better when the agent is more "human like". If Jill cannot answer a question, it provides the learner with guidance (in the form of signifiers) on the closest questions the agent is trained on to help move the conversation into Jill's domain of competence.

A by-product of this project is a schema for writing good syllabi. We now understand the kinds of questions learners ask and have developed an ontology and structure for class syllabus helps find answers. Quite apart from any of consideration of AI, this can help reduce a teacher's load in creating a syllabus, reduce the number of questions students ask because they can easily find the answers in the syllabus, and thus further reduce the load on the teacher because she has to deal with fewer questions. (We will return to this point in Section IV.B.)

## II.C. JILL WATSON SA

The Jill Watson Social Agent (SA) is another part of our AI-powered socio-technical system: it seeks to enhance learner engagement, connect learners with one another, and build micro-communities of learners. This is important because learners in online classes are geographically distributed and the learning is asynchronous, resulting in a lack of learner-learner interactions and a sense of community. In 2016, a Jill Watson Introduction Agent already responded to student introductions on the Piazza discussion forum of the OMSCS KBAI class. In Fall 2019, as students introduce themselves in an online class, the Jill Watson SA uses natural language processing to extract geographic, academic and other information about them, stores them in an encrypted form, responds to the students, and asks them if they would like to opt into community building. If a student opts in, then Jill attempts to connect the student to other students to build micro-communities that enhance learner-learner interactions.

## II.D. AGENT SMITH: BUILDING JILL WATSON Q&A

While the Jill Watson Q&A agent supports both teachers and learners in an online class, developing Jill for a new class can be justified only if it is affordable. This is because building Jill for a new document such as a class syllabus or a users' reference guide requires expertise and can be labor and time consuming. Thus, the emPrize project has developed a novel AI technology called Agent Smith that captures the expertise of our research laboratory in developing Jill agents and makes it efficient for course creators to generate custom Jill agents for their classes. Together Jill and Smith make online education both more achievable and more affordable. (The name Agent Smith derives from the antagonist in the Matrix series of movies who has the capability to clone himself when needed.)

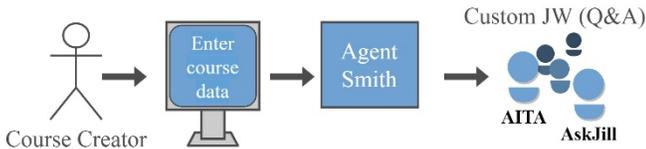

Fig. 2. Agent Smith enables course creators to create their Jill Watson Q&A (2019) agent for answering students' questions in their online classes.

Agent Smith is a novel interactive knowledge-based expert system. Given a new document such as the syllabus for a new class, it starts with the ontology for class syllabi we have developed. The domain of Jill is defined by the ontology and encoded in the form of question templates. Given a course syllabus, Smith builds an episodic memory for events (when is an assignment due?) and a semantic memory for facts and concepts (what is the course late work policy?). Smith next helps the course creator in generating a knowledge base that consists of labeled training questions and answers. It then uses supervised learning to train multiple multiclass classifiers to generate a Jill for the syllabus. The process works much the same way for another document type such as a users' reference guide except that the ontology and the question templates are different.

## III. TECHNICAL RESULTS

### III.A. VERA

As mentioned earlier, VERA is available online (www.vera.gatech.edu), and it uses Encyclopedia of Life (EOL; www.eol.org), the world's largest digital library of biodiversity containing knowledge of more than 1.5 million biological species. EOL's TraitBank supports ecological modeling in VERA in several ways: it provides (i) the ontology of conceptual relations for conceptual modeling, (ii) knowledge of specific interactions among biological species, and (iii) the parameters for setting up the simulations.

Given that the space of simulation parameters can be very large and a learner may not know the "right" values for the parameters, once the learner sets up the conceptual model, EOL uses its knowledge of biological species to directly feed initial values of the simulation parameters into VERA. The learner may then tweak the parameter values and experiment with them. This is another example of learning assistance.

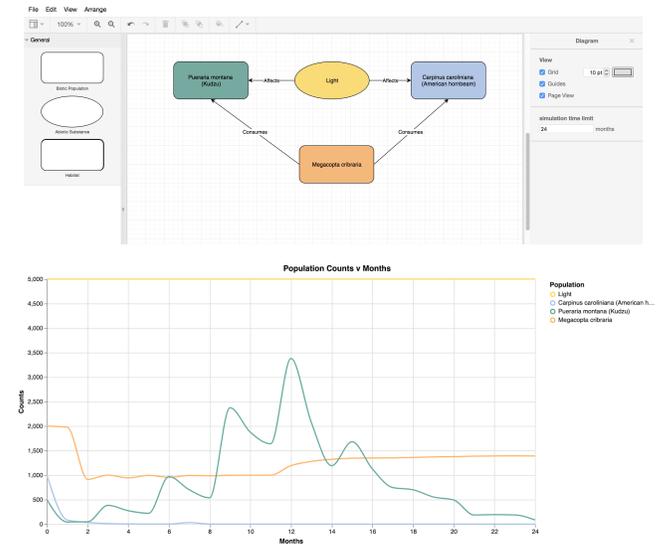

Fig. 3. An example of a conceptual model (the top half of the figure) and its agent-based simulation automatically generated by VERA (the bottom half).

Figure 3 illustrates the use of VERA to model the impact of a kudzu "bug" to moderate the impact of kudzu, an Asian invasive species, on the American hornbeam, a kind of tree common in the eastern half of the United States. In Figure 3(a), the learner interactive builds a conceptual model and in Figure 3(b) VERA illustrates the results of an agent-based simulation of the model. In this case, the simulation results show that because of the introduction of the kudzu bug, the population of kudzu will decline over time and the American hornbeam will survive.

At present, VERA supports modeling of four of the most common types of ecological phenomena: predator-prey, exponential growth, logistic growth, and competitive exclusion. These four types of models together account for a vast majority of ecological phenomena. VERA uses agent-based simulations to provide formative assessment on the conceptual models. Note that VERA automatically spawns agent-based simulations from the conceptual models: an AI agent inside VERA understands enough of the syntax and semantics of both the conceptual models and the agent-based simulations that it can automatically spawn the latter from the former. This is another example of learning assistance in VERA. This learning assistance enables student scientists as well as citizen scientists to model complex phenomena without requiring expertise in the mathematics or the mechanics of agent-based simulations. Further, VERA's support for the whole cycle of model construction, evaluation, and revision fosters self-regulated learning.

We found (An et al. 2019a, 2019b) a strong correlation between the number of hypotheses generated and the complexity of models constructed (r=0.66). Further, use of EOL's domain knowledge helped learners' build more complex models as measured by the number of nodes and links in the models; learners who used EOL frequently were found to come up with more hypotheses and build more complex models (r=0.38; r=0.26). On the other hand, we also found that learners who mastered the process of scientific modeling through formative assessment and self-regulated learning generated more creative models, for example, by hypothesizing novel conceptual relations not specified in the original problem.

### III.B. JILL WATSON Q&A

We have introduced Jill Watson Q&A (2016) and Jill Watson Q&A (2019) agents in multiple online and blended learning classes (as we describe in the next section in detail). Figure 4 shows two examples of students' questions on the Piazza discussion forum of an OMSCS class and Jill's replies to them.

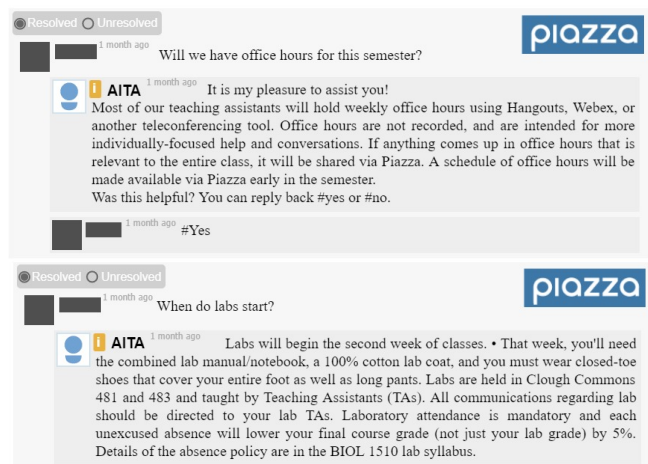

Fig. 4. Two examples of Jill Watson Q&A (2019)'s responses to students' questions in the Piazza forum of an OMSCS class in Fall 2019 (top of figure) and a blended undergraduate class in biology (bottom) in Fall 2019.

Figure 5 illustrates Jill Watson Q&A agents' performance in Georgia Tech OMSCS classes over time. Note that students tend to ask all kinds of questions of Jill, including questions such as "What is the meaning of life?" and "Jill, may I take out on a date?" Clearly, not all of the questions asked of Jill Watson Q&A are valid. Coverage in Figure 5 refers to the percentage of questions that Jill was able to answer in the Piazza forums of the online classes; Precision refers to percentage of questions Jill answered correctly; and Valid refers to the percentage of questions that were in Jill domain of competence. (We did not measure the percentage of Valid questions prior to Summer 2019.)

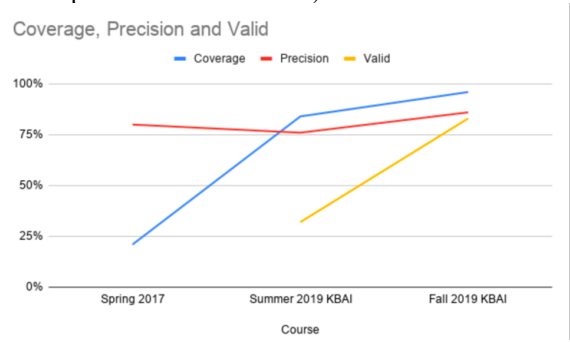

Fig. 5. Improvement in Jill Watson Q&A agent's performance over the course of the emPrize project.

As Figure 5 indicates, Jill's performance has improved quite dramatically over time. Although Jill Watson Q&A (2016) had good precision in Spring 2017 (~80%), its coverage was quite low (~20%). In Fall 2019, Jill Watson Q&A (2019) has a coverage of >96% and a precision of >86%. It is especially noteworthy that the percentage of valid questions has gone up considerably between Summer 2019 and Fall 2019. This is because we now manage learner expectations of AI agents much better so they tend to ask more valid questions. (We will return to this point in detail in Section V.B.)

In addition to class syllabi, Jill Watson Q&A (2019) also answers questions based on the 27-page user reference guide for VERA. Jill for VERA's user reference guide has exactly the same architecture and algorithms as the Jill for class syllabi. The major difference is that the ontology and question templates we have developed for the user reference guide are different from that for class syllabi. Figure 6 illustrates a couple of questions to Jill agent for VERA and Jill's answers to the questions.

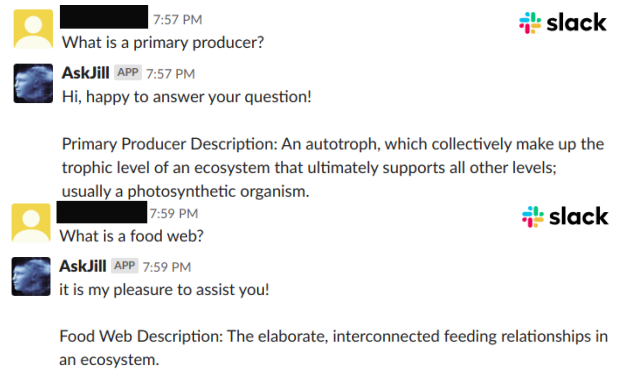

Fig. 6. Two examples of Jill Watson Q&A (2019)'s responses to students' questions on the Slack channel for VERA. Note that while the examples in Figure 4 pertained to the syllabus of a class, these examples relate to domain knowledge.

It is noteworthy that Jill Watson Q&A (2019) runs on both Piazza and Slack (as indicated by Figures 4 and 6). This is important because Slack is quicker and more scalable. It is also worthy of note than while Jill for class syllabi deals almost exclusively with questions about the mechanics of a class, it deals with both mechanics- and content-related questions for VERA's users' reference guide. (We will return to this point in Section V.E.)

### III.C. Jill Watson SA

We have introduced the Jill Watson Social Agent (and before that the Introductory Agent) in multiple online classes (as described in the next section). Figure 7 provides a couple of examples of Jill Social Agent in action on the Piazza discussion forum of an OMSCS class.

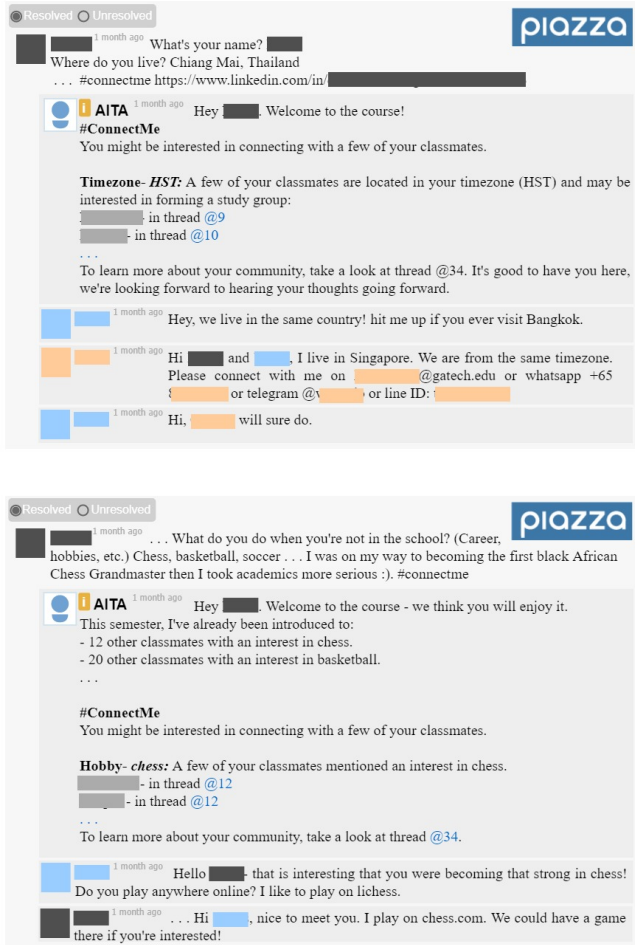

Fig. 7. Two examples of Jill Watson SA's responses to students' introductions. In the first example, Jill helps two students connect based on shared geography; in the second, it connects two students based on a mutual interest in chess.

### III.D. Agent Smith

As indicated above, Agent Smith helps generate Jill Watson Q&A agents for class syllabi as well as VERA's users reference guide. Note that Smith uses the same mechanisms for creating Jill for different types of documents; the difference lies in the different ontologies we have developed for various types of documents. This makes Smith quite efficient in creating a Jill.

## IV. Achieved Problem Impacts

### IV.A. VERA

We have deployed VERA in several educational contexts and at several different locations as summarized in Table 1.

TABLE I. VERA EDUCATION CONTEXT

| Stakeholders | Venue | Dates |
|---|---|---|
| (35) graduate students, residential class, Cognitive Science | Georgia Tech | Spring 2017 |
| (8) citizen scientists | Colorado State University | Summer 2018 |
| (80) undergraduates, blended class, General Ecology | Georgia Tech | Fall 2018 |
| (4) citizen scientists | Citizen Science Conference | Spring 2019 |
| (2) NSF REU students | University of Tennessee, Chattanooga | Summer 2019 |
| (~200) undergraduates, blended class, Introductory Biology | Georgia Tech | Fall 2019 |

As Table 2 indicates, on using VERA, students in the Georgia Tech undergraduate blended class in Fall 2018 showed significant improvement in their knowledge of ecology (An et al. 2019a, 2019b): the average correct response rate in cognitive science class increased from 61.57% on the pre-test to 72.22% on the post-test, whereas the rate in the biology class increased from 71.15% on the pre-test to 80.38% on the post-test. The calculated t-value between the pre-test and the post-test in the biology class was significant (p value=0.0322 at 0.05 level). These results build on similar results we obtained earlier in middle school (Agarwal, Hartman & Goel 2018). It is noteworthy that students in the cognitive science class used VERA for modeling several domains such as economics and epidemiology in addition to ecology. This indicates that architecture of VERA extends to many agent-based domains beyond ecology.

TABLE II. OVERALL RESULTS OF BIOLOGICAL KNOWLEDGE

| Stakeholders | Pre-test | Post-test |
|---|---|---|
| **Cognitive Science class (N=36)** | | |
| Average correct response rate (%) | 61.57% | 72.22% |
| Overall effective (N=36, GLMM, Beta= -0.6389, SD=0.3232, t= -1.977, p= 0.052.) | | |
| **Biology class (N=52)** | | |
| Average correct response rate (%) | 71.15% | 80.38% |
| Overall effective (N=52, GLMM, Beta= -0.4615, SD=0.2125, t= -2.171, p= 0.0322*) | | |

### IV.B. Jill Watson Q&A

Table 3 indicates the deployment of Jill Watson Q&A (2016) and Jill Watson Q&A (2019) over the years. Note that in Fall 2019, Jill is operating in six classes taught by various instructors, including a large online class on Machine Learning (ML4T) with ~1,200 students, a blended class in AI (Blended KBAI) with ~125 students in undergraduate and graduate sections, and a blended undergraduate class in biology (Blended Biology) with ~200 students. This illustrates the growing scope of Jill. Overall, Jill has answered thousands of questions in 13 online and blended classes with >4,000 students over the years. (These numbers do not include Jill for VERA that we had described above.)

TABLE III. JILL WATSON Q&A DEPLOYMENT

We have successfully deployed Jill in 13 online and blended, graduate and undergraduate classes, over the years. This includes 6 classes at present in Fall 2019.

| Stakeholders | Dates |
|---|---|
| Jill Q&A (2016) – OMSCS KBAI | Spring 2016 |
| Jill Q&A (2016) – OMSCS KBAI | Fall 2016 |
| Jill Q&A (2016) – OMSCS KBAI | Spring 2017 |
| Jill Q&A (2016) – OMSCS KBAI | Fall 2017 |
| Jill Q&A (2016) – OMSCS KBAI | Spring 2018 |
| Jill Q&A (2019) – OMSCS KBAI | Summer 2019 |
| Jill Q&A (2019) – OMSCS HCI | Summer 2019 |
| Jill Q&A (2019) – OMSCS KBAI ~ 500 Students | Fall 2019 |
| Jill Q&A (2019) – OMSCS HCI ~ 500 Students | Fall 2019 |
| Jill Q&A (2019) – OMSCS ML4T ~ 1200 Students | Fall 2019 |
| Jill Q&A (2019) – OMSCS EdTech ~ 125 Students | Fall 2019 |
| Jill Q&A (2019) – OMSCS Blended KBAI ~ 125 Students | Fall 2019 |
| Jill Q&A (2019) – OMSCS Blended Biology ~ 200 Students | Fall 2019 |

An important difference between Jill Watson Q&A (2016) and Jill Watson Q&A (2019) is the way we deployed in the online classes. Jill Watson (2016) ran on the general Piazza discussion forum and answered the questions it could; human TAs answered the rest. The goal was to test the authenticity of her answers. However, this also created the possibility of deception. Jill Watson (2019) runs on a dedicated thread on the discussion forum so that the students know they are interacting with a virtual teaching assistant. We made this change in part for ethical reasons - we want the students to know when they are dealing with a virtual TA - and partly for technological reasons - we want Jill Watson to learn from the student feedback on her answers). (We will return to this point in Section VII. on Ethical Evaluations.)

*Reduction in Teaching Load:* Figure 3 in the previous section indicated the significant improvement the Jill Watson Q&A agent over time. We can estimate the approximate time Jill Watson Q&A (2019) saves the teaching staff in answering

TABLE IV. JILL WATSON SA- COMMUNITY BUILDING

Jill Watson SA is presently running the Community-Building feature in four online classes.

| | Total Introductions | Community-Building feature | |
|---|---|---|---|
| | | # Students Opt-in | # Students receive reply |
| OMSCS KBAI | 367 | 166 | 157 |
| OMSCS EdTech | 150 | 68 | 60 |
| OMSCS HCI | 369 | 228 | 215 |
| OMSCS ML4T | 385 | 154 | 142 |

questions. Given a question posted on the Piazza discussion forum of an online class, we estimate it may take a human teaching assistant 1 minute on average to read the question and 2 minutes to type and post the answer; 2 minutes on average to open the course syllabus; and 10 minutes on average to search the answer from the syllabus, for a total of 15 minutes per question. We do not have reliable estimate for the time spent on task switching, i.e. the time taken by a human teaching assistant to switch her attention to the student's question and then switch back to other tasks; we assume 5 minutes per question for task switching, for a total of 20 minutes on average per question.

In the Summer 2019 OMSCS KBAI class, the Jill Watson Q&A (2019) agent answered 136 valid questions. This means the agent saved >45 hours, or >1-person week, of teaching staff's time. Similarly, for the Summer 2019 OMSCS HCI class, where Jill answered 139 valid questions and saved another >45 hours of teachers' time. These numbers are lower than the number of questions students put to Jill in earlier years; of course Jill Watson Q&A (2016) could answer only a much smaller portion of the questions than does Jill Watson Q&A (2019) as illustrated in Figure 3. One likely reason that students now ask fewer questions is because of the improvement in writing the course syllabus that we had discussed in Section II.B: now that we better understand the kinds of questions students ask of a syllabus, and now that we an ontology and a structure for writing syllabi, the class syllabi already tend to contain answers to many student questions and it is also easier for most students to find the answers. (We will return to this point in Section V.B.)

Although the number of valid questions asked of Jill over Summer 2019 was lower than we expected (again please see Section V.B.), we can now estimate the amount of time Jill could save in many more classes or classes with many more students. Given that it correctly answered 136+139=275 questions in two classes with ~1,000 students during Summer 2019 and saved >90 hours of teachers' time, if it was deployed in, say, 20 similar classes (as we are planning for Spring 2020), it may save ~900 hours of teachers' time!

*IV.C. JILL WATSON SA*

In the OMSCS KBAI classes from Fall 2016 through Spring 2018, the Jill Watson (2016) Introductory Agent automatically answered student introductions. In Spring 2017, we found that it automatically generated responses to ~59% of student introductions. However, it did not attempt to further help with learner-learner interactions.

As Table 4 summarizes, in Fall 2019, the Jill Watson SA performed significantly better than the Jill Watson Introductory Agent in Spring 2017. For example, the response rate of Jill Watson SA is now close to 100% and the responses are more varied. Further, Jill Watson SA now helps connect learners in an online class with their peers and form micro-communities as indicated in the previous section. For example, of the 367 students in the OMSCS KBAI class in Fall 2019 who introduced themselves, 166 opted into the community building program; Jill Watson SA was able to generated a community-oriented response for 157 of 166 students (or about ~94%). The qualitative feedback we have received from the online students too is positive.

*IV.D. AGENT SMITH*

As Figure 8 illustrates our estimate of the number of person hours needed to generate a Jill Watson Q&A agent for a new class syllabus. In particular, we estimate that in Spring 2016, it took us ~1500 person hours to generate the first JW. Now that the interactive expert system Agent Smith is operational, we estimate it takes us only about 25 hours of 1 person's work (~25 person hours) to generate a JW for a new class syllabus. This represents a dramatic increase in the efficiency of generating JW. (We will return to this point in the Section V.)

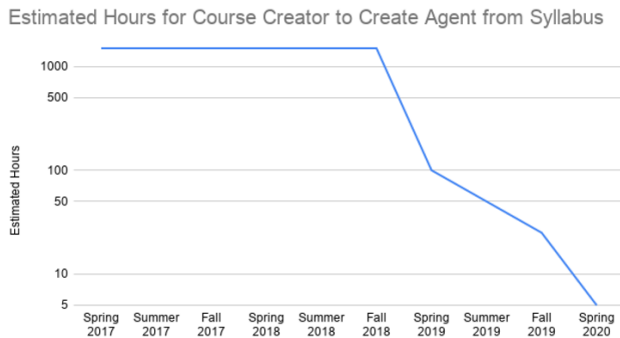

Fig. 8. Agent Smith has helped reduce the amount of time to create a Jill Watson Q&A agent for a new class from >1000 hours to ~25 hours at present. We expect this time to reduce further to ~5 hours by April 2020.

*IV.E. RELATED CONTEXTS AND IMPACTS*

We have performed several kinds of analyses for assessing the learning in the OMSCS KBAI course. First, the completion ratio in the online KBAI class over various semesters is ~80% and thus comparable to the ratio in equivalent residential KBAI course (Goel & Joyner 2017). This is important because the retention ratio in many MOOCs typically is much lower. This indicates that OMSCS students find learning KBAI achievable.

Second, student performance on the learning assessments in the OMSCS KBAI course is comparable to that in the equivalent residential course, where the two sections share the same instructors, syllabus, structure, assessments and graders (Goel & Joyner 2017). Student performance on the learning assessments in the OMSCS KBAI course is also comparable to that in the equivalent residential course using blended learning where the residential students had access to the same video lessons as the online students (Goel 2019). This is important because the quality of learning in many MOOCs has been open to question.

Third, OMSCS students express satisfaction with the online KBAI class including the video lessons and the teaching assistance they receive in the class. For example, Ou et al. (2019) found that a vast majority of online students found the ~150 tutors embedded in the video lessons useful for mastering the subject matter. Further Gonzales & Goel (2019) found that the design of the OMSCS KBAI class fostered self-regulated learning. More importantly, they found that the online students' assessment of teaching assistance in the online KBAI course was high. In fact, students' estimate of learning assistance increased over the term of the course even in the presence the Jill; this suggests that students perceive Jill as providing learning assistance and aiding the learning process.

Fourth, we found that cumulative student activity on the online forum for the OMSCS KBAI class has a positive correlation with student performance in the class as measured by the grades they receive. We also found that a lag in the response to a student's comment on the online discussion forum has a negative correlation with cumulative student activity on the discussion forum. This is noteworthy because if a student asks a question and no human teaching assistant replies for a few hours, then the student may have moved on another issue by the time a teaching assistant does reply. However, if Jill can answer the question promptly, then the student is more likely to be still engaged with the issue. In this way, Jill helps enhance student engagement.

Finally, we surveyed students in the OMSCS KBAI class in Summer 2019 to understand their perceptions and mental models of Jill. We found that while many students find Jill likable and intelligent (at least to some degree), most of them do not consider her to be very "human-like". This is part of the cost of moving from the design of the Jill Watson Q&A (2016) agent that relied on a databank of Q&As and had more authentic replies. To address this problem, we have begun work on adding a personality module to Jill. (We will return to this point in Section VII on Ethical Evaluations.)

## V. EXPECTED PROBLEM IMPACTS

In the first four subsections below, we project ahead only until April 2020 when the XPrize AI competition will end. These projections are based on preliminary results from research already in progress, and thus we can be confident about them. In the fifth subsection, we briefly mention over aspirations over the medium and long terms.

*V.A. VERA*

Three major technological additions to VERA are underway to make it more effective in supporting inquiry-based learning. First, we are adding a library of models of ecological phenomena. We expect this library to act a collaboration space where users can share, critique, copy, and revise one another's models. Second, we are adding the ability to import data from external sources into VERA, along with ML techniques that can recommend modifications to the values of simulation parameters based on the discrepancies between the predictions and the observations. Third, as we mentioned in Section II.A, the architecture of VERA admits modeling of all agent-based domains, not just ecology. Thus, we are using VERA for modeling economic systems. We expect that these additions will make inquiry-based learning using VERA more effective in multiple domains (economics, ecology) in multiple classes (middle school, high school, college) at multiple levels of expertise (student, citizen and professional scientists).

On September 3, 2019, Smithsonian Institution started providing access to VERA directly through the main page on its EOL website (www.eol.org). Now EOL's many users can try out ecological models of several species available in EOL. These species are modeled in VERA using the data directly retrieved from EOL such as lifespan, body mass, offspring count, reproductive maturity, etc. This means that the hundreds of thousands of EOL users across the world, including learners and teachers, as well as citizen and professional scientists now have direct access to VERA. This opens up the potential for online learning in open science. We are now collecting data for analysis. In addition, an CitSci.org, an organization of citizen scientists, is preparing to introduce VERA on its website for free access to citizen scientists engaged in field biology.

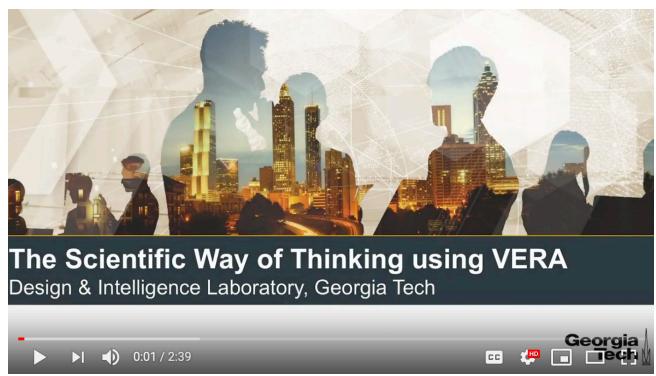

Fig. 9. A screenshot of the first video on "The Scientific Way of Thinking using VERA" on YouTube. The set of videos provides free access to both VERA and Jill Watson Q&A that answers questions about using VERA.

As Figure 9 indicates, on September 16, 2019, we released an online set of videos on "The Scientific Way of Thinking using VERA" on YouTube (https://youtube/Rn5e9fWmPqI). The videos YouTube videos provide access to both VERA and Jill Watson Q&A (2019) for answering questions about using VERA. This opens up the potential for using VERA and Jill for informal online learning in the wild. Again, we are now collecting data for analysis.

*V.B. JILL WATSON Q&A*

In Section IV.B, we noted that in Summer 2019, students seemed to put a smaller number of valid questions to Jill Watson Q&A (2019) than we had expected. Additional experiments in early in the Fall 2019 term reveal the reason. Figure 10 illustrates the relative number of valid questions, the total number of questions asked, and the percentage of valid questions asked. We find that the numbers change quite a bit between two conditions that we have labelled AITA and Sylla. When we introduced Jill to online students as an AI teaching assistant (AITA), they build very high expectations (especially given the hype about AI) and ask all kinds of questions (such as "what is the meaning of life?"). Most of these questions are invalid in that they are outside the class syllabus and Jill is unable to answer them (thus the large number of grey cells on the left of the figure). But when we introduced Jill to the same online students as a Syllabus Agent (Sylla), they ask mostly valid questions and Jill's performance improves dramatically. Thus, in the first weeks of the OMSCS KBAI class this Fall 2019 term, students asked 60 questions of AITA of which only 22 were valid. However, in the same number of following weeks, they asked 80 questions of Sylla of which 72 were valid. We have discovered that an important key to AI powered learning viewed as a socio-technical system is to manage learner expectations of AI agents so that students are more likely to ask questions that are within scope. Now that we have learned this lesson, we expect the total number of valid questions asked of Jill Watson Q&A (2019) agent to increase significantly, perhaps doubling from the 136 questions we had reported in Section IV.B; and higher the number of valid questions, the more are the time savings with Jill Watson Q&A (2019). If this pattern holds through the Fall 2019 term, then the savings to teachers' time would also increase, perhaps as much as doubling to ~90 hours per online class of 500 students.

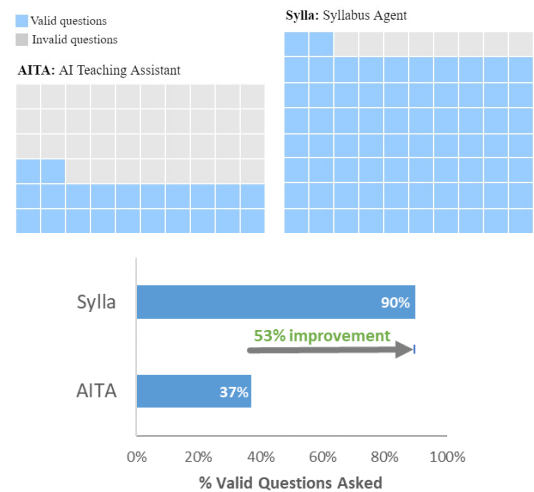

Fig. 10. The number of valid questions students ask and Jill Watson performance in answering them improves when we manage students' expectations by changing Jill Watson's name from AI TA to a syllabus agent (Sylla).

We are presently discussing with the Georgia Tech administration on how to deploy Jill Watson Q&A (2019) in Spring 2020. One proposal is to deploy Jill in all >20 classes in the Georgia Tech Honors Program. This would be a major undertaking because it would cover a variety of disciplines from computing and science, to business and engineering, to humanities and languages. Once Jill successfully works in that many Georgia Tech classes, we hope to offer it to rest of the world engaged in online/blended learning in higher education.

*V.C. Jill Watson SA*

Above we noted the importance of managing students' expectations of the Jill Watson Q&A agent. This is also true of the Jill Watson SA agent. Given that online students take

>10 OMSCS classes over a few years to obtain their degree, some of them have encountered Jill in multiple classes. Figure 11 shows the response of an online student who apparently had encountered the Jill Watson Q&A (2019) in Summer 2019 and now encountered the Jill Watson SA agent for the first time in Fall 2019: this student seems to think that Jill Watson is "getting closer to passing the Turing Test"! (Just to be clear, the emPrize team makes no such claim!)

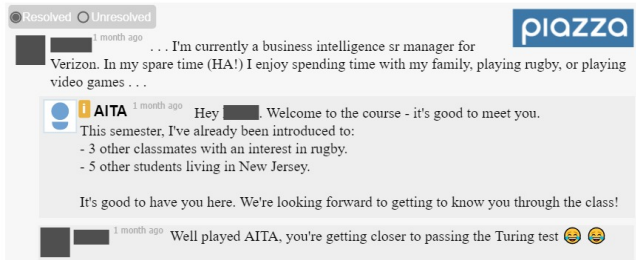

Fig. 11. Students' expectations of AI actors often are very high, all the way up to the level of passing the Turing Test.

We are planning to expand the scope of the Jill Watson SA agent to cover the VERA project: just like the Jill Watson Q&A (2019) agent now answers questions about using VERA based on its 27-page users' reference guide, we want the Jill Watson SA agent to help learner-learner interactions among the VERA users and help build micro-communities based on geography, demographics, academic background, and interests, etc. This is perhaps even more important in case of VERA because VERA is now supporting learning in the wild.

*V.D. Agent Smith*

In Section IV.B, we had performed a rough calculation the amount of teacher' time saved by the Jill Watson Q&A (2019) agent working from a course syllabus. We can do a similar rough calculation for Jill working from VERA's users reference guide. We estimate that Jill may save ~20 minutes per question: ~2 minutes to open up the VERA reference guide, ~15 minutes to locate the answer in the 27-page guide, and ~2 minutes for task switching back to modeling using VERA. Thus, if a user asks 6 questions about using VERA, Jill may save the user 2 hours of time. Further, if a 1000 users ask 6 questions each, it would make for a saving of ~2000 hours. Of course, the exact amount of time saved will depend on the length of the reference guide, the user's familiarity with the guide, and other, similar factors.

Figure 8 in Section IV.D. indicated that we can now use Agent Smith to generate a Jill Watson Q&A (2019) agent for a new class in about 25 hours. This is because Smith captures the expertise our research laboratory has developed in working with JW. We believe that by April 2020 we can bring this number down to ~5 hours for each new class: now that we have captured our laboratory's expertise in building Jill in Smith, we can focus on building the right user interfaces and interactions for using Smith for building Jill. Similarly, for generating Jill for users' reference guides. Figure 11 captures this part of the expected results in the very near future.

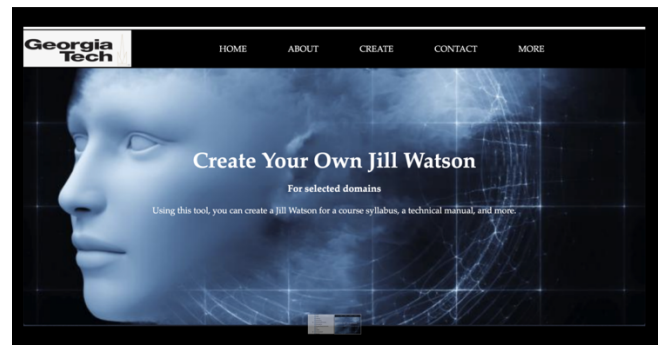

Fig. 12. Part of our vision of the near future. A teacher of a new online course supplies the course syllabus and then works with Agent Smith for ~ 5-6 hours before the class starts. Agent Smith generates a Jill Watson Q&A agent for the class that saves the teacher ~50-100 hours of time answering students' questions during the course.

*V.E. Medium-Term Goals and Long-Term Aspirations*

Our goals for the medium term (5 years) arise directly from this report: (1) A VERA system for all subjects entailing agents (biology, economics, epidemiology, history) and for all levels of education (middle school, high school, college); (2) A Jill Watson Q&A agent for all teachers and students in the world from elementary school to graduate school; (3) A Jill Watson social agent for any learning community, including online and blended, formal and informal learning; (4) An Agent Smith for all types of documents including books; and (5) A complete integration of the VERA and Jill Watson projects.

Our long-term (10-15 years) aspirations also are straightforward but more ambitious. We aspire to completely integrate VERA and Jill Watson along the lines of *A Young Lady's Illustrated Primer* in Neal Stephenson's (1995) *The Diamond* Age in which a young girl with dim prospects for the future comes across an interactive book and is able to learn enough from that in her teens she joins the social elite. We envision that a future VERA will provide simulative social experiences even as a future Jill Watson tells stories about the experiences as in *The Young Lady's Illustrated Primer*. This will directly address our grand challenges: use AI to support learning as a socio-cultural process and make quality education simultaneously accessible, affordable, achievable.

VI. INTRAPROBLEM IMPACTS

*VI.A. General*

Garrison, Anderson & Archer's (1999, 2010) Community of Inquiry (COI) framework is one of the most well-received theoretical frameworks for understanding online learning. COI describes online learning as a socio-technical system in which three elements are critical for quality learning: cognitive presence, teacher presence, and social presence. The emPrize technologies fit the COI framework very well. VERA, for example, fosters self-regulated learning and thus enhances cognitive presence; Jill Watson Q&A amplifies teacher presence by answering student questions based on the syllabus or guide prepared by the teacher; and Jill Watson Q&A and SA together enhance social presence by augmenting engagement and social interactions.

Nass, Fogg and colleagues (Fogg 2003; Nass et al 1997; Nass & Yen 2010) have developed the notion of computers variously as tools, social actors, and media for expression of human interactions. We have repurposed their metaphors for the roles AI technology plays in education. Thus, VERA is a tool for conceptual modeling, but it is also a medium for expressing, sharing and critiquing models of complex phenomena. Jill Watson Q&A is a tool for answering questions, but it is also a social actor for enhancing student engagement. Jill Watson SA is a tool for connecting students, but it is also a social actor for building online learning communities. Agent Smith is a tool; but from Smith's perspective, Jill is a medium for expressing a teacher's answers to questions students may ask. The advantage of these metaphors is that they take us beyond thinking of AI technology only as a tool, which over-simplifies the nature of human-AI interaction and impact; instead, the metaphors of AI technology as a medium and a social actor forces us to consider the socio-technical impact of introducing AI technology to human society in different ways.

*VI.B. VERA*

The notion of virtual laboratories for scientific experimentation is well-established in science learning (De Jong & Van Joolingen 1998). Co-Lab is a collaborative modeling environment where groups of early learners can conduct experimentation via computer simulations to facilitate inquiry-based learning in natural sciences (Van Joolingen et al. 2005); Prometheus is a modeling environment for more advanced modelers that enables them to visualize the structure of models, run them as simulations, and examine their predictions in earth and life sciences (Bridewell et al., 2006). Like Co-Lab, VERA automatically translates user-generated conceptual models into agent-based simulations; Prometheus does not use agent-based simulations. Like Prometheus, VERA uses deep conceptual models with a visual syntax; Co-Lab does have not have deep conceptual models. Unlike both Co-Lab and Prometheus, VERA provides access to a large knowledgebase in the form of EOL thus supporting the cognitive processes of model construction and evaluation that many students find difficult to learn.

*VI.C. Jill Watson Q&A*

Question asking and question answering have received considerable attention in research on intelligent tutoring systems (Graesser et al. 2005) and led to a classification of general types of questions students ask about a subject matter (Graesser, Rus & Cai 2007). However, other than the Jill Watson Q&A (2019) agent, we are not aware of any virtual teaching assistant that automatically answers questions about a class syllabus or a user reference guide as Jill Watson does. The publicly accessible OpenSyllabus project (http://opensyllabusproject.org/) offers access to a large number of class syllabi in a variety of subjects at a range of educational levels; in future, Jill could exploit these publicly available syllabi.

*VI.D. Jill Watson SA*

Prior studies showed that students' sense of community has a strong positive correlation with perceived cognitive learning, perceived learning engagement, class satisfaction, and learning outcomes (Liu, 2007). In addition, there has been more recent research on chatrooms for peer-to-peer learning in large online classes. The evidence from this research is mixed. While Coetzee et al. (2014) found that peer-to-peer assessment in chatrooms was not effective in supporting learning outcomes, Kulkarni et al. (2015) suggest pedagogical strategies for making collaboration in large online classes more effective. Our experience with online learning aligns more closely with the more positive view (Joyner, Isbell & Goel 2016). However, we are unaware of any AI technology for enhancing social interactions by connecting students to form micro-communities as Jill Watson SA does.

*VI.E. Agent Smith*

We are unaware of any other AI technology that interactively builds question-answering teaching assistants for online learning. Insofar as we know, Agent Smith is unique in its capabilities.

## VII. ETHICAL EVALUATIONS

In Section II.B we noted that while Jill Watson Q&A (2016) answered questions based on a databank of previous Q&A pairs in a class, Jill Watson Q&A (2019) answers questions based on the class syllabus. There were two main reasons for this change: technological and ethical. From a technological perspective, as we described in our 2018 XPrize report, by 2017 we had reached a plateau in the percentage of questions Jill Watson Q&A (2016) answered correctly. Here we describe the ethical considerations that led to the change to Jill Watson Q&A (2019). First, when we initially deployed Jill in an online class in Spring 2016, Jill operated in the general discussion forum as if it was just another teaching assistant. In fact, we did not tell the students that Jill was a virtual teaching assistant and the students did not recognize Jill as such either. While we did this experiment with IRB approval to examine if Jill could pass the test of authenticity, and while no student who has interacted with Jill has ever complained to us about it, nevertheless, and as we discuss in Eicher, Polepeddi & Goel (2018), this experiment contained a degree of deception. As described in III.B above, Jill Watson Q&A (2019) operates on a dedicated thread that explicitly notes that it is a virtual teaching assistant.

Second, in analyzing Jill Watson (2016) Introduction Agent's behavior that we briefly mentioned in Section II.C, in 2017 we found preliminary evidence of gender bias. When a male student would say "Soon, I will become father for the first time", the Jill Watson (2016) Introduction Agent may respond with "Congratulations on the impending arrival!" However, when a female student would say "I am pregnant and I am due later this semester", the Jill Watson (2016) Introduction Agent may reply with "Welcome to the class!" without noting the impending arrival of a baby. As we discuss in Eicher, Polepeddi & Goel (2018) in detail, our analysis indicated that this gender bias arose because of the skewed student demographics: ~85% of the students in the OMSCS program at that time were male. As a result, ~85% of the questions/introductions in the class were originated by male students as well. This meant that the Jill Watson (2016) Introductory Agent had a richer databank for answering questions raised by male students than by female students; hence, the gender bias. Of course, this was absolutely unacceptable; hence, the change from Jill Watson Q&A (2016) that answered questions/introductions based on databanks to Jill Q&A Watson (2019) that answers questions based on class syllabi.

## VIII. CLOSING STATEMENT

We started this report by stating two grand challenges for the emPrize project: use of AI to simultaneously address the issues of accessibility, affordability, achievability and quality of online learning in higher education, and creation of an AI-powered socio-technical system for supporting online learning as cognitive and socio-cultural processes. Let us examine our progress in addressing these two goals.

First, let us consider the latter goal. The Jill Watson Q&A agent is an AI tool that reduces the load on the teacher, amplifies the reach of the teacher, assists learners with their questions anywhere anytime, and enhances learner engagement by answering their questions promptly; the Jill Watson Social Agent is an AI actor that helps learners form micro-communities and thus increases social presence in online learning; and Agent Smith helps create a Jill Q&A agent for new online classes, and views Jill as a medium for expressing a teacher's answers to learners' questions. Similarly, VERA is medium for expressing and sharing learner's model of a complex phenomenon, and also a tool for simulating the model, thereby providing prompt formative assessment; further, it supports self-regulated learning, thus enhancing cognitive presence in online learning.

In regard to the first (and larger) goal, (1) the VERA project has established not only that it enhances the quality of learning in scientific domains such as ecology, but also that it simultaneously improves accessibility, affordability and achievability of science learning; for example, VERA is now available online and accessible through both EOL and YouTube; (2) the Jill Watson Q&A project has established not only that it can help reduce teachers' workload while amplifying their reach, but also that it simultaneously provides learning assistance to learners in online classes and enhances their engagement, thus making learning more achievable – Jill Watson Q&A too is now available through Slack and YouTube; (3) the Jill Watson SA project has established that it can enhance learner-learner interactions and help form learning micro-communities, thus enhancing achievability; and (4) the Agent Smith project has established it not only can help generate Jill Watson Q&A agents for a variety of educational documents from class syllabi to user reference guides, but also that it can do so efficiently enough to reduce the overall cost of answering questions and thus contributing to making education more affordable.

In a special issue of The Chronicle of Higher Education on "The Digital Era: how 50 years of information age have transformed education forever", Myers & Lusk (2016) called virtual assistants exemplified by Jill Watson one of the most transformative educational technologies over the last 50 years. The emPrize project is beginning to fulfill that promise.

## IX. IRB STATEMENT

We have IRB approval for conducting all studies with human subjects described above. This include studies in the College of Computing OMSCS classes (Knowledge-Based AI, Human-Computer Interaction, Machine Learning for Trading. and Educational Technology) as well as the blended KBAI class and the Introduction to Programming class. This also includes classes in the School of Biological Sciences (Introduction to Biology, and General Ecology). In addition, it includes experiments in our laboratory described above. We append the IRB certificates in the Appendix to this report.

# Addendum to AI-Powered Learning

emPrize Team, Design & Intelligence Laboratory, Georgia Institute of Technology; emprize@cc.gatech.edu

*Abstract* — The emPrize project continues to make rapid progress towards (1) the design of AI-powered sociotechnical systems for online learning for higher education, (2) scalability and generalizability of our AI technologies for assisting both teachers and learners, (3) development of novel AI techniques for interactive training of the AI virtual assistants, and (4) measurement of the impact of the AI-powered sociotechnical system on online learning in higher education.

## I. PROBLEM AREA UPDATE

United Nations' Sustainable Development Goals (SGDs) include Quality Education (goal #4). The SDGs' also include Reduce Inequalities (goal #10). Thus, the first grand challenge for the emPrize project is to "create and use AI technology for making quality education simultaneously more accessible, affordable and achievable" (Goel 2019), thereby making it more equitable. However, quality education requires both learning assistance and social interaction. Thus, our second grand challenge is to "design socio-technical systems that include AI tutors but also engage social interaction and learning assistance" (Goel 2019). The specific goal of the emPrize project also remains the same (Goel 2019): address the two grand challenges in the context of online learning in higher education.

While the grand challenges and specific goals have remained the same, our framing of the problem has evolved in two ways. First, the integrative framework of Community of Inquiry (CoI; Garrison, Anderson & Archer 1999; 2010) provides a promising (but heretofore unrealized) strategy for designing sociotechnical systems for online learning. CoI advocates virtual learning environments that enhance *cognitive presence*, *teacher presence*, and *social presence*. Cognitive presence pertains to active and experiential learning that motivates and engages the learner situated in a virtual learning environment; teacher presence attends to participation by the teacher in the learning; and social presence deals with social interactions among learners. In our work, a virtual learning environment may refer to an online classroom or an online educational resource. This explains three of the four core AI technologies emPrize has developed (Goel 2019): (1) VERA, the virtual research assistant that supports online experimentation using Encyclopedia of Life, enhances cognitive presence in using an open-source online educational resource; (2) Jill Watson Q&A, the virtual teaching assistant that automatically answers learners questions in Georgia Tech's OMSCS classes, enhances teacher presence in the online classes; and (3) Jill Watson SA, the virtual social assistant that connects learners in OMSCS classes to form micro-communities, together with Jill Q&A enhances social presence in the same classes.

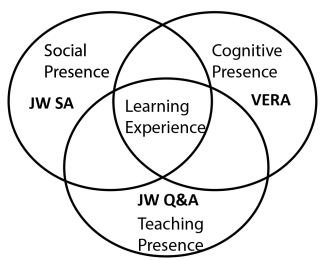

Second, personalization, scalability and generalizability of AI technologies are critical for their use in practice. As just one example, use of AI tools typically requires time, effort, and expertise in AI. However, teachers typically are both very busy and very autonomous: they want easy-to-use AI tools for developing AI assistants in their own manner. This explains the fourth core AI technology emPrize has developed: (4) Agent Smith that empowers a human teacher to interactively develop a Jill Watson Q&A assistant for her class in her own fashion and within the span of several hours. This both personalizes the development of Jill and scales its use. We have used Agent Smith not only for developing Jill for Georgia Tech OMSCS classes, but also for answering questions on the internet based on VERA's User's Guide, thereby both generalizing Jill and completing the integration of the four technologies.

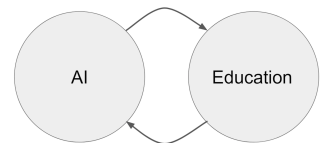

## II. TECHNICAL SOLUTION UPDATE

Since the beginning of the Georgia Tech Fall 2019 semester in late August, we have made two significant technical advances. First, we have enhanced the learning component in Jill Watson Q&A. After a teacher creates a Jill for her online course, Jill starts automatically answering questions from the learners in the class. If Jill cannot answer a question or is not confident of its answer, it forwards the question to the teacher along with recommended answers. The teacher chooses the best answer and the agent sends the best answer to the student while learning the classification of the new question. Jill can also learn directly from the students by reading their evaluation of its answer to a question. After each answer from Jill, students are automatically asked for feedback (#yes for useful answer, or #no). On the next pass, Jill reads this response and either updates the strength of the classification (#yes) or forwards the original question to the course creator (#no). In this way, Jill uses interactive, incremental, supervised ML to continuously improve its performance.

Second, VERA now has the capability of both importing external data that captures the "ground truth" and graphically comparing the results of a simulation with the observed data. We are now using ML for inductive process modeling so that, given a discrepancy between results of a simulation and observed data, VERA can recommend how to tweak the simulation parameters to better match the data.

## III. TECHNICAL RESULTS UPDATE

We have continued to extend, deploy and evaluate Jill Watson Q&A in multiple classes. Since the Fall 2019 semester, we have used Jill in 8 Georgia Tech classes taught by different professors in different academic units, , including graduate online classes in computing, undergraduate blended classes in biology, and, professional education classes in systems engineering.

| Semester | Course | Enrollment | Agent |
|---|---|---|---|
| Spring 2020 | ISYE6501 - 1 | 803 | JWQA |
| Spring 2020 | ISYE6501 - 2 | 317 | JWQA |
| Spring 2020 | KBAI | 566 | JWQA/SA |
| Spring 2020 | ML4T | 731 | JWQA/SA |
| Spring 2020 | EdTech | 253 | JWQA/SA |
| Spring 2020 | HCI | 457 | JWQA/SA |
| Fall 2019 | BIOL-1510 | 229 | QA |
| Fall 2019 | KBAI | 130 | JWQA/SA |
| Fall 2019 | KBAI | 685 | JWQA/SA |
| Fall 2019 | EdTech | 286 | QA/SA |
| Fall 2019 | ML4T | 1209 | QA/SA |
| Fall 2019 | HCI | 647 | QA/SA |

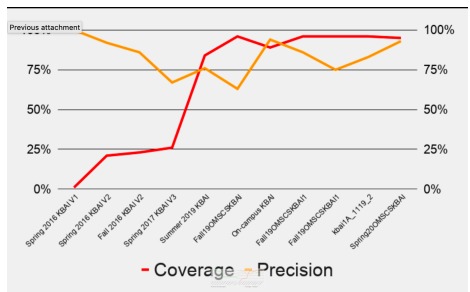

The figure above illustrates that Jill's coverage of valid questions and precision of answers has steadily increased over time; the coverage and precision presently stand at 95% and 92%, respectively. The figure on the left indicates that the time taken to construct a Jill for a new class has steadily decreased. Agent Smith now enables construction of a Jill in <10 person hours of work.

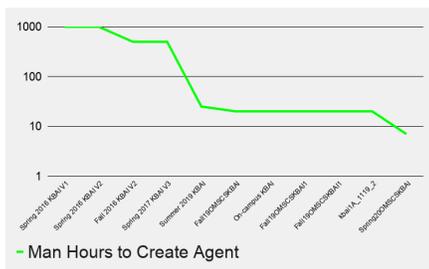

## IV. ACHIEVED PROBLEM IMPACTS UPDATE

We introduced VERA into a Georgia Tech blended course on introduction to biology in Fall 2019. The intervention was conducted during one class period for 50 minutes to 142 students. This version of VERA also contained Jill Watson Q&A for answering questions about VERA's User Guide. We found that both VERA and Jill worked together flawlessly. This is important because it demonstrates the integration of our AI technologies.

Through Smithsonian Institution's Encyclopedia of Life, VERA has attracted a growing international userbase. In January 2020 alone, VERA attracted over 200 site visitors from over 40 countries. This growing dataset is invaluable in understanding how users build and experiment with models.

*Jill Watson Q&A:* We conducted longitudinal surveys of students to understand their mental models of Jill Watson Q&A. Overall, we found that the students were positive about the virtual teaching assistant. A little surprisingly, we found that students' perception of Jill change significantly in terms of perceived intelligence and "human-likeness" over time. This suggest a need for Jill to have flexible characteristics to cater to students' changing perceptions of the agent.

*Jill Watson SA:* We also conducted surveys of students to gather student feedback of Jill Watson SA. Again, overall, online learners were positive about Jill. We found that online learners utilized the information provided by Jill to form connections through shared identities, i.e., students with similar background, interests, etc. Highlighting shared identities among online learners also fostered a sense of belongingness in the class.

## V. EXPECTED IMPACTS UPDATE

We can envision impacts of our AI technologies over several time horizons. In the immediate future (through the end of the XPrize AI competition in April 2020), we expect to continue extending and integrating our technologies. For example, we are expanding the scope of Jill Watson SA to cover the VERA research assistant alongside Jill Watson QA. We are moving towards increased user collaboration within VERA by using Jill to provide recommendations about other users and models. Using EOL lookup data and user modeling behavior, Jill can build connections between VERA's global user base by highlighting shared model interests. Ultimately, JWSA on VERA will be used to drive collaborative model editing and the creation of global communities of citizen scientists.

In the near term, through the end of the calendar year 2020, we are negotiating three kinds of deployments of our technologies within Georgia Tech: (1) deployment of Jill Watson Q&A in all Georgia Tech online classes (2) deployment of Jill in all student dormitories at Georgia Tech, and (3) deployment of VERA, Jill Watson Q&A and Jill Watson SA in Georgia Tech micro-campuses across the country and around the world.

Due to advances in computing and AI, workforce upskilling and reskilling is becoming a major problem in the US and in the world as a whole: workers who lose jobs because of automation need to be retrained for new roles and responsibilities. Workforce development increasingly engages online education: while we may expect a 12year old child to go to school each day and an 18-year old to leave home for college for a few years, for learning workers and working learners, we must take education to where they live and work. Given the large demand, workforce development engages also learning at scale, with tens and hundreds of thousands of learners in an online class or millions of learners using an online educational resource. Thus, in the medium-term of 5-10 years, each year we expect to address at least a million learners taking online classes. Similarly, for learners using online educational resources such as Encyclopedia of Life, we intend to engage a million learners each year. Overall, we expect to engage a few million learners over the next few yeas and several million over the next ten. by the 5$^{th}$ year of the project. Between learners in online classes and learners using online educational resources, we plan to study and support 2-3 million users over the 5 years of the project.

## VI. INTRAPROBLEM IMPACT

We already have alluded to several impacts on AI itself. For example, the development of a technique for training virtual teaching assistants such as Jill Watson Q&A is a contribution to both machine learning and human learning.

We have made a noteworthy impact in a non-obvious and unexpected direction. Since Fall 2014, we have collected data on tens of thousands of questions asked by thousands of learners in a score of AI classes in Georgia Tech OMSCS, blended and residential educational programs in computing. We have also collected data on answers given by the teachers and teaching assistants in these classes. In addition, since Spring 2016, we have collected data on Jill Watson Q&A's answers to some of these questions. In particular, since Spring 2019, we have collected data on Jill's answers based on the class syllabi. As a result of analyzing all this data, we have developed a novel schema for writing good class syllabi so that the teacher specifies all the required information and a learner can easily find the needed information. In practice, this leads to fewer questions, which saves time for both teachers and learners.

This teacher-learner (human-human) interaction is part of our AI-powered sociotechnical system for online learning in which humans and AI agents live, work and learn together. For example, a teacher may write a preliminary syllabus for his online class. The teacher may then use Agent Smith to create a Jill Watson Q&A agent to automatically answer questions based on the class syllabus. As an expert system in question answering, Agent Smith presents templates of questions asked by learners in previous classes ands thus guides the teacher to write a good syllabus. Next, Jill answers learners' questions in the online class. If it succeeds in answering a question, it seeks feedback from the students and learns from the experience; it is fails, it asks the teacher for the correct answer and again learns from it. At the end of the course, the teacher can examine all the questions learners asked and the answers Jill Watson gave. At any point during this cycle, the teacher can use Agent Smith to modify Jill as desired.

## VII. CLOSING STATEMENT

We started by positing two grand challenges for online learning in higher education. How much progress has emPrize made towards addresssing them?

*Design of AI-powered sociotechnical system for learning asssitance:* Both VERA and Jill Watson Q&A provide learning assistance, in the context of online educational resources on the internet where learners may engage in self-directed learning, and Jill in online classses that are part of an educational program. VERA is now being used on the internet from users across the world. For example, a biology class in Costa Rica apparently has adopted VERA as an interactive modeling tool. In the near future, we expect VERA as becoming the modeling tool of choice for the 1 million visitors to EOL each year. Jill Watson Q&A has been used in 17 Georgia Tech classes, including gradute online classes in computing, undergraduate blended classes in biology, and, this term, also professional education classes in systems engineering. By now >5,000 learners as well as >100 teachers and teaching assistants have worked with Jill Watson.

*Design of AI-powered sociotechnical system for sociol interaction:* The Jill Watson SA helps form micro-communities by connecting geograhpically disgtributed learners engaged in asynchronous learning. Jill Watson Q&A also helps learner engagement. Immediacy of feedback, whether from teachers, peers or AI assistants appears to be correlated to higher engagement, and better engagement seems correlated with superior performance.

*Quality of Learning:* Quasi-experimental studies indicate that online learners in our AI classes have the same performance on learning assessments and the same completion ratio as residential students in equivalent face-to-face classes. This result holds even when adjusted for student demograophics (Goel 2019a). Surveys of students in online AI classes indicate that their evaluation of teaching assistance in the class increases through the semester even when they know of the presence of an AI teaching assistant (Gonzales & Goel 2019).

*Accessiblity, Affordability, and Achievabaility of Learning:* VERA enables learners to conduct virtual experiments for free using public domain resources and thereby ground their knowledge in data. Jill Watson Q&A amplifies the reach and capabilities of teachers in online classrooms. It offloads some of the teaching tasks to AI assistants and frees up precious time for more creative engagements with learners. Agent Smith enables a teacher to efficiently and easily build her own Jill in a manner of her choice. Within five years, we would like a Jill for every teacher in the world. "We have received so much interest in our technologies that we decided to spawn a startup named "Beyond Question".

The emPrize Team

Our team's name "emprize" means "an adventurous enterprise".

The emPrize team is composed of faculty, staff and students associated with Georgia Tech's Design & Intelligence Laboratory. Current members include Varsha Achar, Sungeun An, William Broniec, Ida Camacho, Ashok Goel, Marissa Gonzales, Eric Gregori, David Joyner, Rohit Mujumdar, Brady Young, and Qiaosi Wang.

Former members of the team include Parul Awasthy, Pramodith Ballapuram, Fredrick Bane, Robert Bates, Christopher Cassion, Bobbie Eicher, Siddharth Gulati, Kaylin Hagopian, Sung Jae Hong, Joshua Killingsworth, Yaroslav Litvak, Swati Mardia, Keith McGreggor, Heather Newman, Sydni Peterson, Lalith Polepeddi, Ananya Roy, Angela Smiley and Shimin Zhang.

PI Goel is solely responsible for the contents of this report.

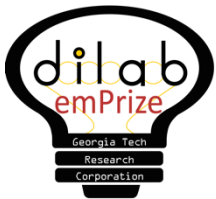

Acknowledgements

The emPrize team is sponsored by the Georgia Tech Research Corporation. It has received funding and support from Georgia Tech, Georgia Tech's College of Computing, School of Interactive Computing, School of Computational Science and Engineering, Georgia Tech's Center for 21st Century Universities, and Georgia Tech Professional Education.

The project on the Virtual Research Assistant for Computational Experimentation (VERA) is supported through a NSF South BigData Spoke grant, an NSF PPSR grant, and an NSF REU grant.

Our research partners have included Smithsonian Institution's Encyclopedia of Life, Citsci.org at Colorado State University, NSF's I-Corps program, NSF South BigData Hub, and Georgia Tech's Online MS in CS program.

We are especially grateful to Dr. Jennifer Hammock (Project Director, Encyclopedia of Life, National Museum of Natural History, Smithsonian Institution), Dr. Gregory Newman (Colorado State University), Dr. Hong Qin (University of Tennessee at Chattanooga), Dr. Spencer Rugaber (Georgia Tech School of Computer Science), and Dr. Emily Weigel (Georgia Tech School of Biological Sciences).

We thank Smithsonian Institution for granting access to its EOL digital library and software, and IBM for access to its Bluemix platform.